# A Categorization Scheme for Socialbot Attacks In Online Social Networks

**Silvia Mitter**
Knowledge Technologies Institute
Graz University of Technology,
Austria
smitter@student.tugraz.at

**Claudia Wagner**
Institute for Information and
Communication Technologies
JOANNEUM RESEARCH Graz,
Austria
claudia.wagner@joanneum.at

**Markus Strohmaier**
Knowledge Technologies Institute
Graz University of Technology,
Austria
markus.strohmaier@tugraz.at

**Abstract**
In the past, online social networks (OSN) like Facebook and Twitter became powerful instruments for communication and networking. Unfortunately, they have also become a welcome target for socialbot attacks. Therefore, a deep understanding of the nature of such attacks is important to protect the Eco-System of OSNs. In this extended abstract we propose a categorization scheme of social bot attacks that aims at providing an overview of the state of the art of techniques in this emerging field. Finally, we demonstrate the usefulness of our categorization scheme by characterizing recent socialbot attacks according to our categorization scheme.

**ACM Classification Keywords**
K.4.2 Computing Milieux - Computers and Society: Social Issues

**Introduction**
Online social networks (OSNs) like Facebook and Twitter can be used for many different purposes, such as searching advice or organizing political protests. Such usage can be disturbed arbitrarily, as for instance shown in [16] where researchers report that certain hashtags on Twitter, which were used to organize protests regarding the Russian parliamentary election, were spammed until they become useless as a channel of communication among protesters. Recently *socialbots*, which are automated agents in OSNs that can perform certain actions on their own have spread in OSNs. Past research for example inspected a variety of characteristics for distinguishing between users susceptible and non-susceptible to socialbot attacks [18]. In this work, we extend this research and propose a comprehensive categorization scheme of social bot attacks. We demonstrate the utility of the categorization scheme by categorizing exemplary OSN attacks, described in previous research.

**Categorization of Socialbot Attacks**

To gain a deeper understanding of the variety of socialbot attacks, we developed a categorization scheme of socialbot attacks to aid future comparison and systematic investigations into this emerging phenomenon. Our categorization scheme allows categorizing social bot attacks along different dimensions. In the following we describe our methodology, and then proceed to introducing the categorization scheme in detail. Finally, we show how the categorization scheme can be used to categorize exemplary socialbot attacks which recently took place in OSNs.

**Methodology**

The categorization scheme was created following an iterative process, by executing the following steps:

- We inspected existing research literature to identify similarities and differences between socialbot attacks as basis to form dimensions and categories within dimensions.

- We considered the *Six problems of categorization* introduced in [14].

- Then we evaluated the quality of our categorization system by considering properties and requirements based on previous work regarding categorization schemes (e.g. [3][4][8][11][13])[1]: u*nambiguous and well defined, repeatable and objective*, *exhaustive*, *useful*, *based on technical details,* identifying *internal vs. external threats* and *similar but multiple classification*. Since attacks may be complex we *do not claim mutual exclusiveness.*

- We demonstrate applicability of the scheme by categorizing existing attacks (described in publications or on media sites), as shown in Table 1.

- Finally, we discussed our proposal with other researchers, and incorporated feedback.

**A Categorization Scheme for Socialbot Attacks**

Based on several former taxonomy proposals in related fields (e.g.[9]) we defined dimensions for describing and categorizing socialbot attacks: *Targets*, *Account Types*, *Vulnerabilities*, *Attack Methods* and *Results*. Every dimension is hierarchically built as a tree where leafs represent the categories. Dimensions differ in width and depth, but the levels of abstraction within dimensions are consistent. In the following we explain the utility of each dimension in detail.

---

[1] This evaluation was not performed in a strict manner, rather we used these requirements during the development of our categorization scheme to guide and steer the process.

*Targets*

One or several targets are involved in socialbot attacks. Therefore we identify *who* or *what* is attacked as a target. This is not necessarily the same as who or what comes to harm in the end (see *Results* dimension). As shown in Figure 1, an individual entity, a collection of entities, or the OSN itself can be the target of an attack. We define an entity as a resource, which has its own space within the OSN. Examples can be users, events or locations, which have their own pages within the OSN, such as user accounts or hashtags on Twitter. A collection of entities can be formed by social relations, by traits (rather static properties of an entity such as the user account age of an user entity) or by states (rather dynamic properties such as the user mood or interest of a user entity or the number of attendees of an event entity).

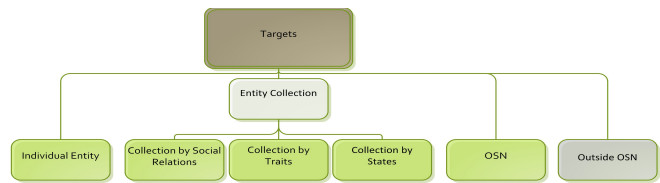

*Figure 1: Targets Dimension*

*Vulnerabilities*

Attacks usually require certain vulnerabilities (see Figure 2), which can be exploited in order to facilitate the attack. We distinguish between OSN-specific and general system vulnerabilities which have already been analyzed and categorized in previous work (e.g. see the CVE[2] standard) and are therefore not further discussed here. Vulnerabilities described in this categorization scheme focus on specific OSN functionalities caused by the system or by users. System vulnerabilities can emerge from the tradeoff between providing an unrestricted, uncensored platform and the need for security. For example unprotected *Entity Information* can enable potential attackers to retrieve valuable information about a user's relations and activities. *IP Usage* vulnerabilities allow attackers to send requests with no or insufficient IP limitation. Also vulnerabilities regarding *API Usage or Unverified Account Creation* can arise. *User Account Action*s describe missing system regulations regarding how many actions are allowed per user within a given time frame. For example, ineffective CAPTCHAs can enable adversaries to obtain access to the system with no or little effort. Unrestricted *Channel or Topic Usag*e means that everyone can use communication channels in an unrestricted manner - e.g. one does not have to belong to a specific community in order to use the community's communication channel.

---

[2] http://cve.mitre.org/

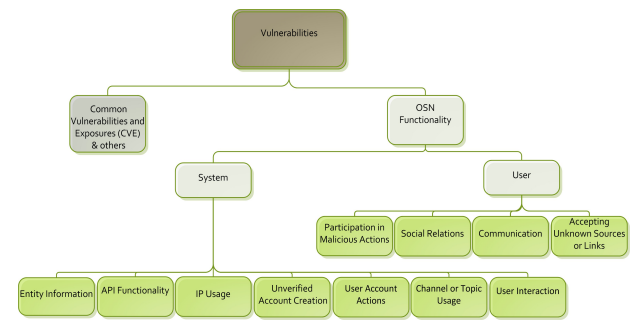

*Figure 2: Vulnerabilities Dimension*

Unrestricted *User Interactions* allow users to interact with each other regardless of their social relations or other information and restrictions. Since OSNs are driven by users and their behavior, vulnerabilities also arise from user behavior. Users may *participate in*

*malicious actions*, participate in an untrusted *communication*, establish social connections with strangers (untrusted *social relations*) or *accept unknown sources or links*.

*Attack Methods*
Different socialbots adopt different attack methods (see Figure 3). An attack method describes *how* an attack is conducted. Again, we concentrate on attack methods within the OSN functionality. OSN *topics* can be used in an *abusive way*, changing the initial meaning of a topic to a specific new one (*hijacking* a topic) or just destroying the meaning (*censoring* a topic) by automatically creating a large amount of new entries. *Unsolicited Communication* means that socialbots send messages and communicate in an unsolicited way. *Clickjacking* attacks try to trick users to click on links embedded in unobtrusive context. It can be used within the abusive usage of topics or within unsolicited communication, and therefore clickjacking itself is not represented as a separate attack method in our categorization scheme. The same applies to *Affiliation Attacks*, where attackers try to make a user buy something on a specific website and in return receive commission. Another attack method is *Spoofing* which is defined as the action of impersonating a specific user to perform an attack.

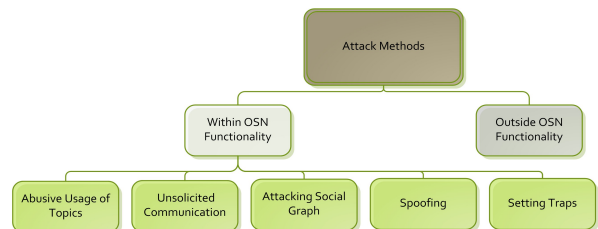

*Figure 3: Attack Methods Dimension*

*Setting Traps* may be a rather passive attack method. It can be used to e.g., attract specific users interested in specific topics or malicious accounts e.g., by using bots in a similar way as *Honeypots*. Honeypots are known from the security sector, where they are used to send adversaries down the wrong track to distract them from real potential system vulnerabilities. In OSNs they may arouse interest without directly addressing other accounts. Another example of setting traps is to leave traces. In some OSNs users can e.g. see who visited their profile and socialbots may perform an attack by leaving traces to gain users' interest.

*Account Types*
We distinguish between four different account types (see Figure 4). *Compromised Accounts* originally belong to benign users but are taken over by socialbots. *Creepers Accounts* belong to users willing to e.g., temporarily sell their accounts for advertising purposes via services such as *pay4tweet*[3] or *Pay with a Tweet*[4]. Accounts, only created for being used within attacks are called *Fraudulent Accounts*. Finally, if the OSN allows it, socialbots may also conduct attacks within the OSN without using any account. Categorizing attacks originating from outside the OSN (exogenous to the system) are beyond the scope of this work.

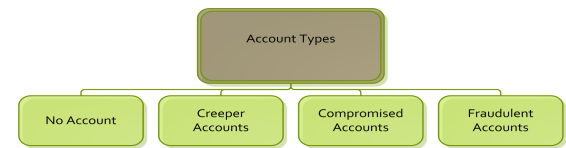

*Figure 4: Account Types Dimension*

---

[3] http://www.pay4tweet.com/

[4] http://www.paywithatweet.com/

*Results*

Finally, different socialbot attacks lead to different results (see Figure 5), which are the observable outcome of an attack. We split results depending on whether an active change in the OSN is achieved or not. *Changed Social Graph, Modified Communication Channel or Topics, Injured Accounts* (*hacked or blocked*) and *Injured OSN* (e.g. long response time) are active results. A passive result is if an attacker gains access to sensible data.

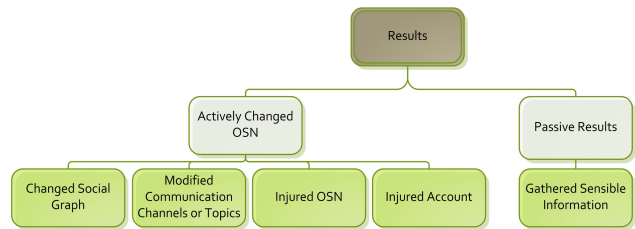

*Figure 5: Results Dimension*

## Applying Categorization Scheme to Attacks

In Table 1 we exemplarily show how our categorization scheme can be used to describe and categorize attacks in OSNs. Towards this end, we use previous research regarding Spam or bot attacks in OSNs. Since the main focus of existing research often concentrates on defense mechanisms etc, attacks are often described only partially and therefore cannot be categorized along all dimensions.

## Conclusions

In this work we present a categorization scheme for social bot attacks which provides a much needed first overview of the state of the art of attacks in OSNs. We hope that our work represents a stepping stone for more principled investigations into the role of socialbots in OSNs. We believe that a better understanding of such attacks is essential in ensuring that OSNs become a trustworthy and effective tool for exchanging of ideas and information.

| Dimension | Category | | | Empirical Example |
|---|---|---|---|---|
| Target | Indidividual Entity | | | [2,5] |
| | Entity Collection | by Social Relations | | [5,15] |
| | | by Traits | | no example |
| | | by States | | [1,16] |
| | OSN | | | no example |
| | Outside OSN | | | out of scope |
| Account Type | No Account | | | no example |
| | Creeper | | | *pay4tweet, paywithtweet* |
| | Compromised | | | [6,7] |
| | Fraudulent | | | [1,2,5,12,15,16,17] |
| Vulnerability | CVE and others | | | out of scope |
| | OSN Functionality | System | Entity Information | [1,2,5,15] |
| | | | API Functionality | [1,5,7,12,15,16,17] |
| | | | IP Usage | [16] |
| | | | Unverified Account Creation | [1,2,5,12,15,16,17] |
| | | | User Account Actions | [5,17] |
| | | | Channel or Topic Usage | [6,7,10,12,16,17] |
| | | | User Interaction | [1,5,6,7,15,17] |
| | | User | Participation in Malicious Actions | *pay4tweet, paywithtweet* |
| | | | Social Relations | [1,2,5,12,15] |
| | | | Communication | [1,2,15] |
| | | | Accepting Unkown Sources or Links | [7] |
| Attack Method | Within OSN Functionality | Abusive Usage of Topics | | [7,10,12,16,17] |
| | | Unsolicited Communication | | [1,6,7,15,17] |
| | | Attacking Social Graph | | [1,5,15] |
| | | Spoofing | | no example |
| | | Setting Traps | | [2,12] |
| | Outside OSN Functionality | | | out of scope |
| Result | Actively Changed OSN | Changed Social Grpah | | [1,2,5,12,15,17] |
| | | Modified Communication Channels or Topics | | [1,2,6,7,10,12, 15,16,17] |
| | | Injured OSN | | no example |
| | | Injured Account | | no example |
| | Passive Results | | | [5] |

*Table 1: Applying categorization scheme to socialbot attacks*

**Acknowledgements**

This work was supported in part by a DOC-fForte fellowship of the Austrian Academy of Science to Claudia Wagner.